\documentstyle[secnumtab]{fbssuppl}
\input psfig.sty
\title{Final State Interactions in Near Threshold Meson Production from 
pp Collisions\thanks{Dedicated to Prof. A. Weiguny on occasion 
of his retirement}\thanks{Supported by  FZ J\"ulich}}
\author{A. Sibirtsev, W. Cassing}
\institute{Institut f\"ur Theoretische Physik, Universit\"at Giessen \\
D-35392 Giessen, Germany}
\begin{document}
\maketitle
\begin{abstract}
We analyse the experimental data on near threshold $\pi$, $\eta$, 
$\omega$,
$\eta'$ and $K^+$ production from $pp$ collisions and show that all 
information gained so far is compatible with  approximately constant 
production matrix elements when including the rescattering between 
the baryons in the final states. Different methods to include the 
final state interactions are discussed and their range of validity
is indicated. We, furthermore, show that Dalitz plots for the proton-meson
invariant mass spectra at different energies should be suited to 
distinguish between final state interactions and resonant production 
amplitudes.  
\end{abstract}

\section{Introduction}
The strong interaction at  low energies, i.e.  
elastic nucleon-nucleon scattering, is reasonably described by 
$\pi$, $\sigma$, $\eta$, $\rho$ and $\omega$-meson
exchanges between the nucleons and the detailed experimental data on 
$NN{\to}NN$ reactions have provided information about the 
meson-nucleon-nucleon vertices, i.e.  coupling constants and  
form factors. 
Above the pion production threshold the dominant inelasticy of 
the $NN$ interactions is due to pion production. 
Already in 1960 Woodruff~\cite{Woodruff} 
proposed to extend the  $NN$ potential model in order to
calculate $NN{\to}NN{\pi}$ reactions. 
Near the reaction threshold the contribution from $\Delta$ intermediate 
states is expected to be negligible and the $S$-wave pion production 
is governed by the ${\pi}NN$ vertex. Thus pion production  
is suited to verify our knowledge about the ${\pi}NN$ coupling constant.

A similar motivation also holds for near threshold $\eta$-meson
production, when the $S_{11}$ resonance replaces 
the $\Delta$, and the $\eta$ production cross section should provide
some information about the ${\eta}NN$ vertex. Note that the  
status of the ${\eta}NN$ coupling constant is still an open 
problem~\cite{Tiator,Feuster,Sibirtsev1} since within our present
knowledge  $g_{{\eta}NN}$ might vary between 1 and 9  depending 
on the model adopted as well as the accuracy of the  experimental data.

Near threshold $\omega$, $\phi$ and $\eta^\prime$-meson 
production in $NN$ collisions should provide information about 
the relevant $MNN$ couplings as well as  on  
intermediate baryonic resonances that might be 
coupled strongly to these mesons; this is discussed as $hidden$ 
resonance properties. Obviously the strangeness production in  
$NN$ collisions involves an additional mechanism due to  
strange meson exchange ($K, K^*$) and sheds light on
the kaon-hyperon-nucleon vertex. 

We will base our analysis in this work on the combined efforts of many 
experimental groups that have taken data on near threshold meson
production: These type of experiments for
$NN$ collisions were started at the Indiana University Cyclotron
Facility with data on the $pp{\to}pp\pi^0$ reaction
at excess energies $\epsilon{=}\sqrt{s}{-}2m_N{-}m_\pi$ from 
$\simeq$1 to $\simeq$30~MeV~\cite{Meyer1,Meyer2}. 
The data at $\epsilon{\leq}$1~MeV were complemented  by 
CELSIUS (Uppsala)~\cite{Bondar}; recently also IUCF 
reported~\cite{Hardie,Flammang} new cross sections on the
$pp{\to}pn\pi^+$ reaction at $\epsilon{<}$20~MeV. 
The near threshold $\eta$-meson production in $pp$ collisions
was studied at SATURNE by the collaborations SATURNE-II~\cite{Bergdolt}
and PINOT~\cite{Chiavassa} and at
CELSIUS~\cite{Calen1}. These measurements cover the range
${\simeq}1.5{\le}\epsilon{\le}100$~MeV. In 1998 CELSIUS reported 
also data~\cite{Calen2} on the $pn{\to}pn\eta$ reaction at 
$16{\le}\epsilon{\le}100$~MeV.
Additionally, the $pp{\to}pp\eta^\prime$ reaction was studied at SATURNE
by SPES-III~\cite{Hibou} and at the COoler SYnchrotron (J\"ulich)
by COSY-11~\cite{Moscal} at $\epsilon{<}$10~MeV. 

Furthermore, the $pp{\to}pp\omega$ reaction
was measured at SATURNE by the  DISTO Collaboration;
they reported~\cite{Balestra} data 
on $\omega$-meson production from $pp$ collisions at 
$\epsilon{\simeq}320$~MeV and $\phi$ production at 
$\epsilon{\simeq}82$~MeV. The data on the $pp{\to}pp\omega$ 
reaction at $\epsilon{<}$31~MeV were measured by SPESIII 
and have been reported only very recently~\cite{Hibou1}.
The $pp{\to}p{\Lambda}K^+$ reaction was measured
by the COSY-11~\cite{Balewski} and the COSY-TOF~\cite{Bilger}
Collaborations; COSY-11 also has reported on the 
$pp{\to}p\Sigma^0K^+$ reaction~\cite{Sewerin}.

It should be noted that
apart from the $\pi^0$ and $\eta$ data the experimental
results on near threshold meson production in $NN$ collisions
have became available only during the last years. This has initialized
a lot of theoretical activity and inspired the most
recent calculations within meson-exchange models. 
Here we present a systematical analysis
of the data and provide the relation between the 
experimental observables and the production mechanism, respectively.

Our work is organized as follows: In Section 2 we will describe the 
threshold kinematics and discuss various approaches for the final 
state interactions (FSI). Section 3 is devoted to an analysis of 
the available data with the aim to extract average production 
matrix elements for the mesons measured so far. In Section 4 we will 
discuss the effect of FSI and resonance amplitudes on differential 
observables while Section 5 concludes this study with a summary.

\begin{figure}[t]
\label{memo9}
\centerline{
\psfig{file=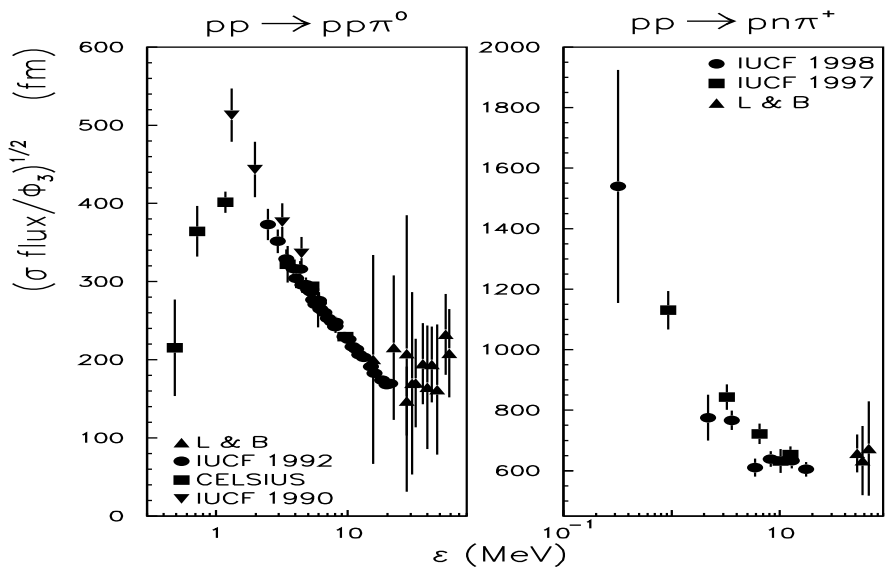,height=8cm,width=13.5cm}}
\phantom{aa}\vspace{-0.8cm}
\caption[]{Experimental data on the 
$pp{\to}pp\pi^0$ \protect\cite{Meyer1,Meyer2,Bondar,LB}
and $pp{\to}pn\pi^+$ \protect\cite{Hardie,Flammang,LB}
reaction amplitude as a function of the excess energy $\epsilon$.}
\phantom{aa}\vspace{-0.7cm}
\end{figure}

\section{Threshold kinematics and Final State Interactions}
The threshold kinematics have several features that request
specific conditions for the experimental measurements as well as  
the theoretical analysis. We note that within nonrelativistic 
approaches the 
three-body phase space $\Phi_3$ is proportional to $\epsilon^2$.
Thus data on threshold meson production are frequently analyzed in
terms of the $reduced$ $cross$ $section$  $\sigma{/}\epsilon^2$.
In~\cite{Sibirtsev2} we have proposed to analyze the data in a more 
transparent way in terms of an average reaction amplitude 
(for fixed invariant energy $\sqrt{s}$) as
\begin{eqnarray}
\label{average}
|M_R| &=&
2^4 \ \pi^{3/2} \  \lambda^{1/4}(s,m_N^2,m_N^2) \ \sqrt{\sigma s}
\nonumber \\
&\times& \left\lbrack \ \intop^{(\sqrt{s}-m_a)^2}_{(m_b+m_c)^2}
\hspace{-0.3cm} \lambda^{1/2}(s,s_1,m_a^2) \ 
\lambda^{1/2}(s_1,m_b^2,m_c^2) \ 
\frac{ds_1}{s_1} \right\rbrack^{-1/2}
\end{eqnarray}
with ${\lambda}(x,y,z){=}(x{-}y{-}z)^2{-}4yz$ and $m_a, m_b, m_c$ denoting the
masses of the particles in the final state.

Furthermore, among the five variables characterizing the three-body 
final state, there are two of direct physical relevance:  the invariant
mass $\sqrt{s_1}$ of two final particles $b$ and $c$  and the 
4-momentum squared $t$ transfered from the initial nucleon to
particle $a$. These variables allow to express the production amplitude  
in the meson-exchange mechanism. Note that $\sqrt{s_1}$ varies from 
$m_b{+}m_c$ up to $m_b{+}m_c{+}\epsilon$. Since the width 
of the known baryonic resonances is larger than 100~MeV, 
it is not possible - within a  narrow $\epsilon$ range - 
to detect directly an intermediate baryonic resonance coupled to 
$bc$ (meson + nucleon or hyperon) and to reconstruct experimentally 
the relevant production mechanism~\cite{Sibirtsev1,Sibirtsev3}.
Therefore complete measurements have to be performed at least up 
to $\epsilon{\simeq}$100~MeV.

Close to  threshold both $\sqrt{s_1}$ and $t$ vary only slightly 
and the production amplitude itself is expected to be almost 
constant. Fig. \ref{memo9} shows the amplitudes for  the
$pp{\to}pp\pi^0$ and $pp{\to}pn\pi^+$ reactions extracted from the
experimental data~\cite{Meyer1,Meyer2,Bondar,Hardie,Flammang,LB}
using Eq. (\ref{average}). The amplitudes substantially depend
on the excess energy $\epsilon$ but seem to approach a constant 
value for large excess energies. Such a deviation from a constant 
value has been predicted by Watson~\cite{Watson} and 
Migdal~\cite{Migdal} due to the strong $S$-wave interaction between 
the final nucleons.

\begin{figure}[h]
\phantom{aa}\vspace{-0.4cm}
\centerline{
\begin{minipage}[l]{6cm}
\psfig{file=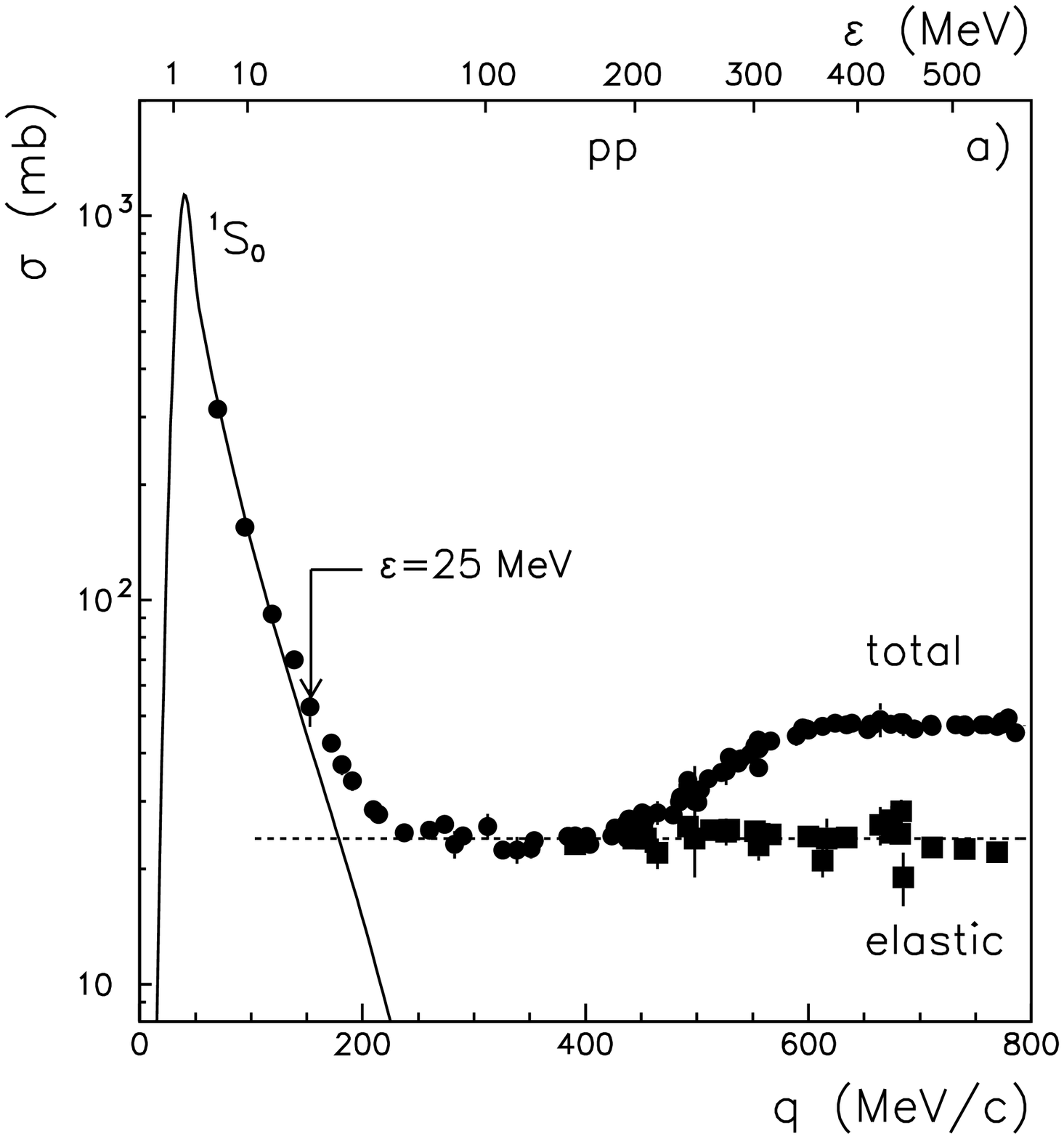,width=6.2cm,height=8.6cm}
\end{minipage}\begin{minipage}[l]{6cm}
\psfig{file=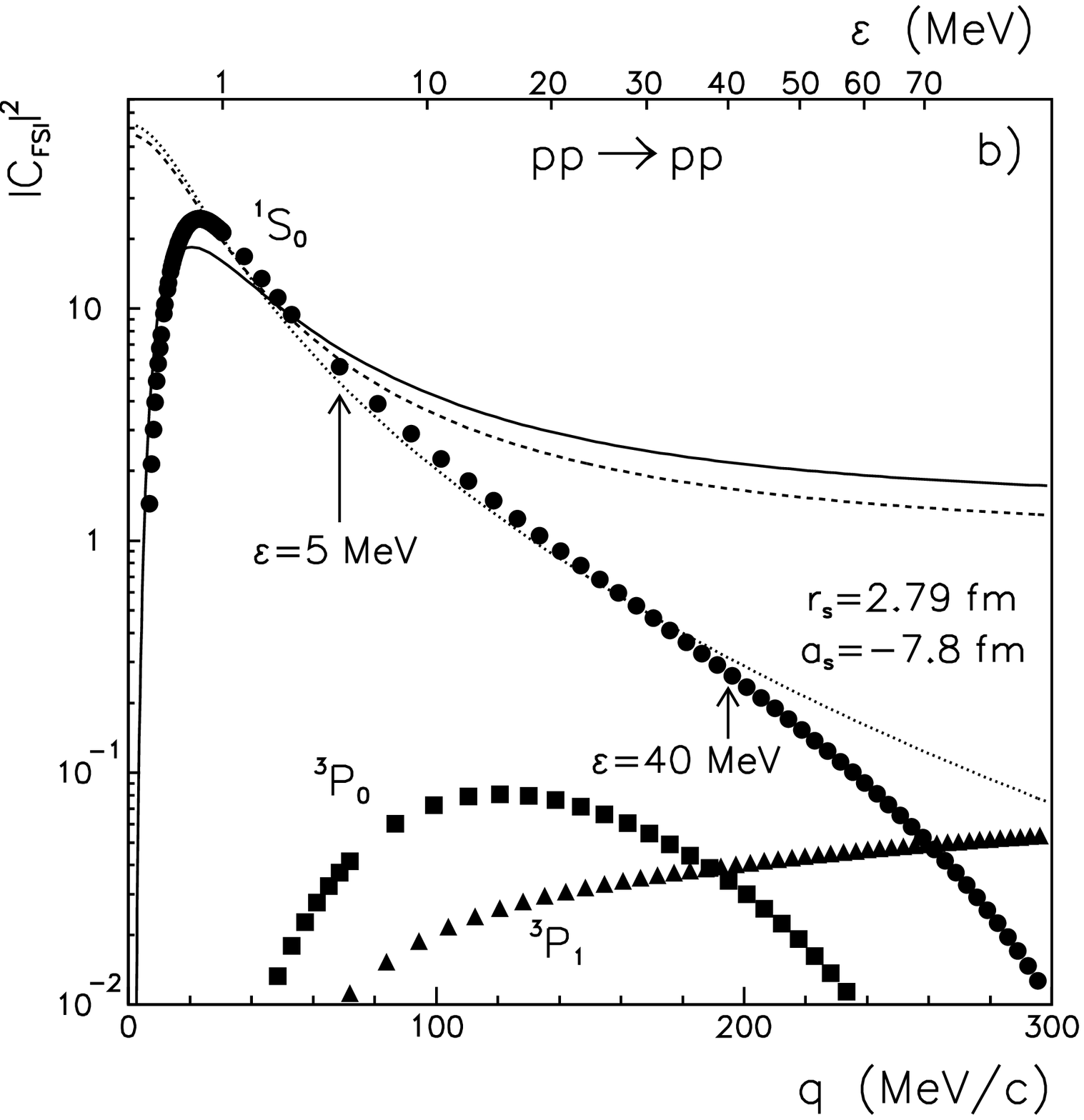,width=6.2cm,height=8.6cm}
\end{minipage}}
\phantom{aa}\vspace{-0.6cm}
\caption[]{a) Total (circles) and elastic (squares) cross
sections for the $pp$ interaction as a function of the momentum in the
$pp$ cms. The data are from Ref. \protect\cite{LB}.
The solid line shows the contribution from the $^1S_0$ partial 
wave~\protect\cite{SP99}, while the dashed line indicates the  
large momentum limit. b) Correction factor due to 
final-state-interactions (FSI). 
The squared  $pp$ scattering amplitudes are shown for $^1S_0$ (circles), 
$^3P_0$ (squares) and $^3P_1$ (triangles) partial 
waves~\protect\cite{Nijmegen}. The dotted line shows the result from
the effective range approximation, the dashed line shows the 
inverse squared 
Jost function without Coulomb correction, while the solid line includes a
Coulomb correction. Further notations are explained in the text.}
\label{memo6-8}
\end{figure}

Indeed the Watson-Migdal theorem can be understood, for instance,
in terms of the $pp$ cross section shown in Fig.\ref{memo6-8}a) as
a function of the proton momentum $q$ in the center-of-mass
system. The cross section is enhanced at low $q$  due to 
the $^1S_0$ partial wave~\cite{SP99} as shown by the
solid line in Fig. \ref{memo6-8}a). Above about 400 MeV/c  
the elastic cross section approaches again a constant as 
indicated by the dashed line. It is thus expected that the 
production of mesons is enhanced when the protons
emerge with a low relative momentum in the final state.

Note that in the $NN{\to}NNM$ reaction the momentum $q$ varies from 
zero up to ${\simeq}\sqrt{m_N\epsilon}$. Obviously at large 
excess energies the contribution from FSI due to the strong 
$S$-wave to the total $pp{\to}pp\pi^0$ cross section seems to be
not dominant, since one should integrate over the 
wide phase space. However, $S$-wave FSI can be detected by
differential observables even at large $\epsilon$ as
we will illustrate in the following. For $\epsilon{\le}25$~MeV
the FSI between the protons is entirely due to the $^1S_0$-wave.
At higher energies the $pp$ cross section deviates from the
calculations with the $^1S_0$ phase shift as  can be seen from 
Fig.\ref{memo6-8}a) and is indicated by an arrow.

Now the deviation of the $NN{\to}NN\pi$ reaction amplitude 
shown in Fig.\ref{memo9} from a constant
can be understood within the Watson-Migdal approximation. Moreover, the
difference in the energy dependence of the $pp{\to}pp\pi^0$
and $pp{\to}pn\pi^+$ reaction amplitudes at $\epsilon{\le}1$~MeV
is due to the Coulomb interaction between the final nucleons, 
which is absent in
$np$ scattering, but influences the final $pp$ scattering
for $\epsilon{\le}1$~MeV as can be seen from Fig.\ref{memo6-8}b.

Taking the near threshold production amplitude 
$M$ as  a  constant, it was proposed in Refs.\cite{Watson,Migdal} 
to factorize the reaction amplitude $M_R$ as
\begin{equation}
\label{wm}
M_R = M \times C_{FSI} ,
\end{equation}
where $C_{FSI}$ stands for the amplitude due to the interaction 
between the final particles. Strictly one should account 
for the three-body FSI, which itself is a rigorous problem.
As was suggested by Gell-Mann and Watson~\cite{GellMann} the 
near threshold $NN{\to}NN\pi$ reaction might be examined when 
considering the dominance of low energy $NN$ scattering as 
compared to the $S$-wave ${\pi}N$  interaction and taking 
$C_{FSI}$ as the $S$-wave $NN$ on-shell scattering amplitude $T_s$. 
Obviously,  the produced particles are off-shell before  
rescattering due to FSI, which in principle involves an additional
assumption about the off-shell correction to $T_s$.

Fig.\ref{memo6-8}b) shows the squared $^1S_0$, $^3P_0$ and 
$^3P_1$ $pp$ scattering amplitudes  calculated with 
the phase shifts from the Nijmegen partial wave 
analysis~\cite{Nijmegen}. At low $\epsilon$ the $S$-wave 
amplitude dominates  and for further 
implementation to $NN{\to}NNM$ calculations can be  
expressed within the effective range approximation as
\begin{equation}
\label{scat}
T_s(q) = \left( -\frac{1}{a_s}+\frac{r_s q^2}{2}-iq \right)^{-1},
\end{equation}
where $a_s{=}{-}7.8$~fm and $r_s{=}2.79$~fm~\cite{Stoks} denote the
scattering length and  effective range, respectively. 
The effective range approximation is shown in Fig.\ref{memo6-8}b) 
by the dotted line and is valid for excess energies $\epsilon$  from 
1 up to 40~MeV.

Another way~\cite{Taylor} to account for FSI is to 
express $C_{FSI}$ as an inverse $S$-wave Jost function  
\begin{equation}
\label{Jost}
C_{FSI}(q) = \frac{q+i\beta}{q-i\alpha},
\end{equation}
where the parameters $\alpha$ and $\beta$ are related to the
effective range parameters as
\begin{equation}
\label{range}
a_s=\frac{\alpha + \beta}{\alpha \beta}, \hspace{2cm}
r_s= \frac{2}{\alpha + \beta}.
\end{equation}
The squared inverse Jost function is shown by the dashed 
line in Fig.\ref{memo6-8}b) and is close to the effective range 
approximation only for $\epsilon{\le}5$~MeV. Note that Eq.(\ref{Jost})
approaches unity at large momenta $q$ since the $S$-wave FSI
does not contribute at large $q$, which is the proper boundary
condition in terms of the factorization (\ref{wm}). Furthermore, to 
account for the Coulomb repulsion at $\epsilon{\le}1$ one can 
correct $C_{FSI}$ in line with the Gamov factor (solid line in Fig. 2b).

Finally, when  calculating  the FSI within different approaches as
the $NN$ scattering amplitude itself or with the Jost function or an 
effective range approximation including Coulomb corrections we 
find no severe differences up to excess energies of 
${\simeq}5$~MeV. Furthermore, since the $S$-wave dominates the 
$NN$ scattering up to $\epsilon{\simeq}25$~MeV, the Jost function 
is an appropriate way to account for FSI corrections because
it approaches unity at large $\epsilon$ in line with the
factorization ansatz. The disadvantage of the method is due
to the implementation of the on-shell $NN{\to}NN$ amplitude. 
However, off-shell corrections will introduce new parameters 
to the calculations that later on should be controlled by data.

\begin{figure}[t]
\psfig{file=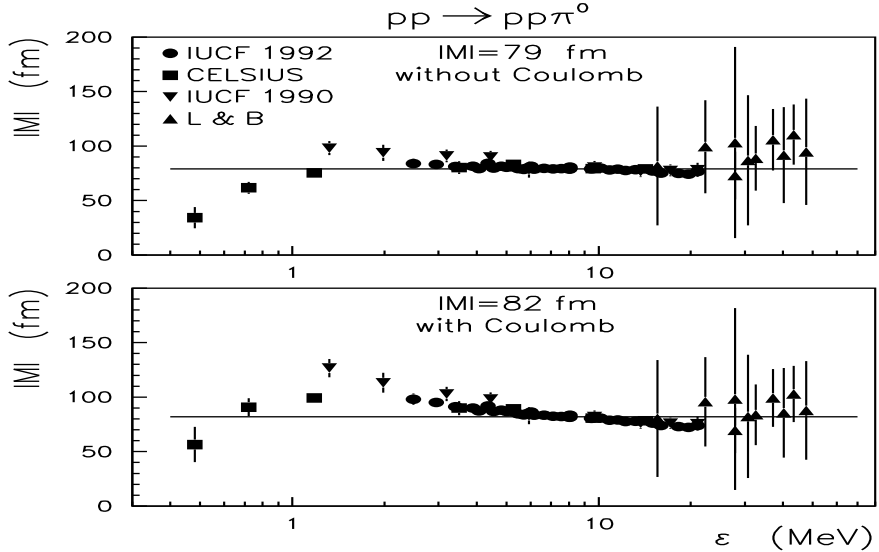,width=13.0cm,height=8cm}
\phantom{aa}\vspace{-1.1cm}
\caption[]{Experimental data~\protect\cite{Meyer1,Meyer2,Bondar,LB}  
on the average $pp{\to}pp\pi^0$ production amplitude as a function of the
excess energy $\epsilon$ calculated without (upper part) 
and with Coulomb correction (lower part).
The solid lines show the fit with a constant value for the 
production matrix element.}
\label{memo3}
\phantom{aa}\vspace{-0.8cm}
\end{figure}

\section{Evaluation of the production amplitude from the data}
Now we adopt the on-shell approach and use the Jost function in 
order to account for the FSI correction. Moreover, we perform the data
analysis with and without Coulomb correction to demonstrate the
systematic uncertainties. To calculate the
production amplitude $|M|$ we substitute the function $C_{FSI}(q)$ 
in the integral of Eq.(\ref{average}). 

Fig.\ref{memo3} shows the  average $pp{\to}pp\pi^0$ production 
amplitude as a function of the excess energy $\epsilon$. 
In this representation the data are almost energy
independent and approach a constant value. For $\epsilon{<}1$~MeV 
two data points from Ref.\cite{Bondar} substantially deviate from 
the constant for calculations without the $pp$ Coulomb repulsion, 
but become closer to a  constant value after Coulomb correction. 
However, to shed light on the Coulomb effect one needs more data 
at $\epsilon{<}1$~MeV.
We also notice that the 1992 IUCF data~\cite{Meyer2} are better described 
by a constant amplitude $|M|$ as compared to the 1990 IUCF 
data~\cite{Meyer1}. Our analysis with Coulomb correction
gives $|M|\approx 82$~fm for the $pp{\to}pp\pi^0$ reaction while we get
$|M| \approx$ 79 fm without this correction which indicates the 
systematic uncertainty of our analysis.
 
\begin{figure}[h]
\phantom{aa}\vspace{-0.8cm}
\centerline{
\begin{minipage}[l]{6cm}
\psfig{file=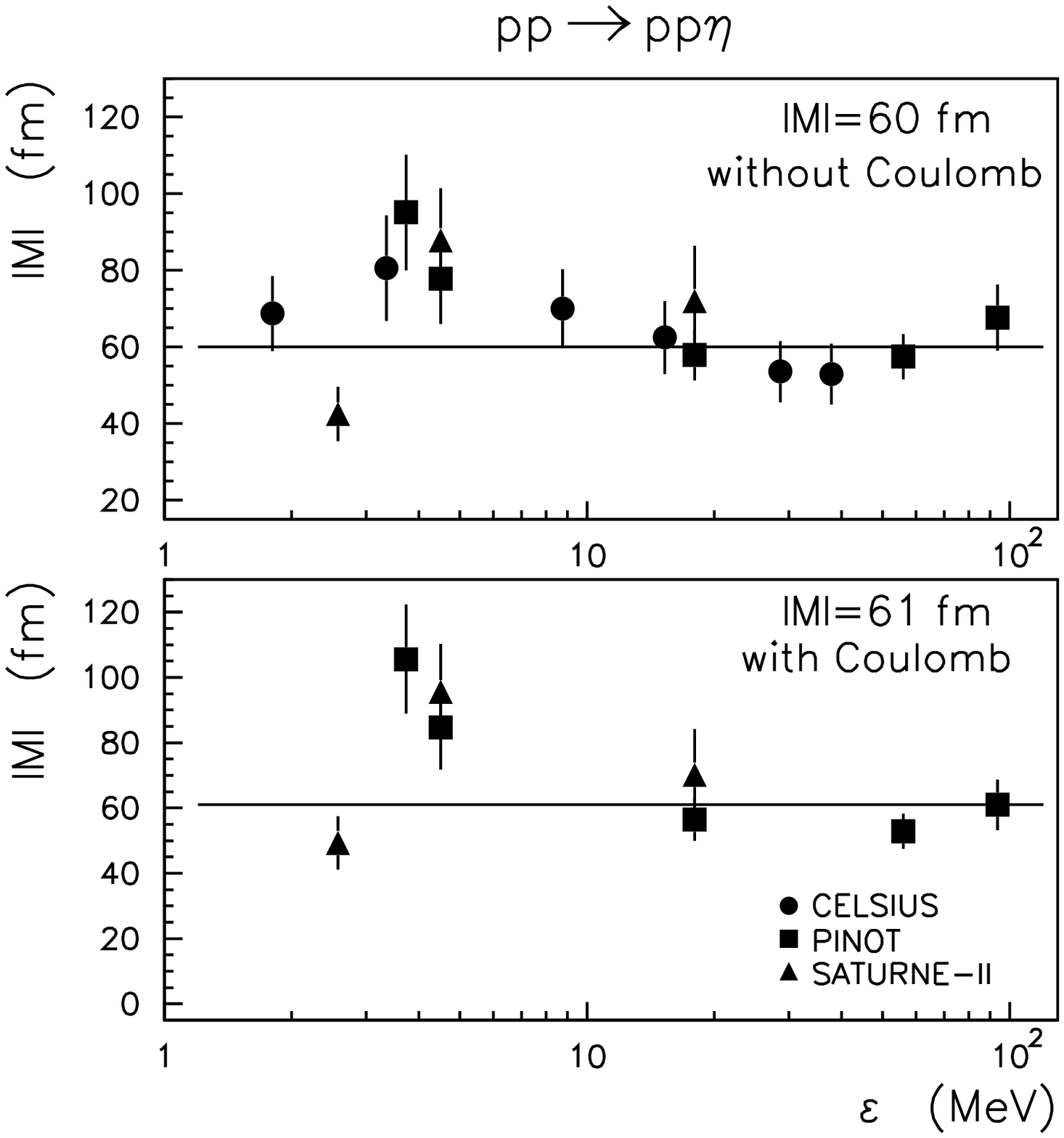,width=6.2cm,height=8cm}
\end{minipage}\begin{minipage}[l]{6cm}
\psfig{file=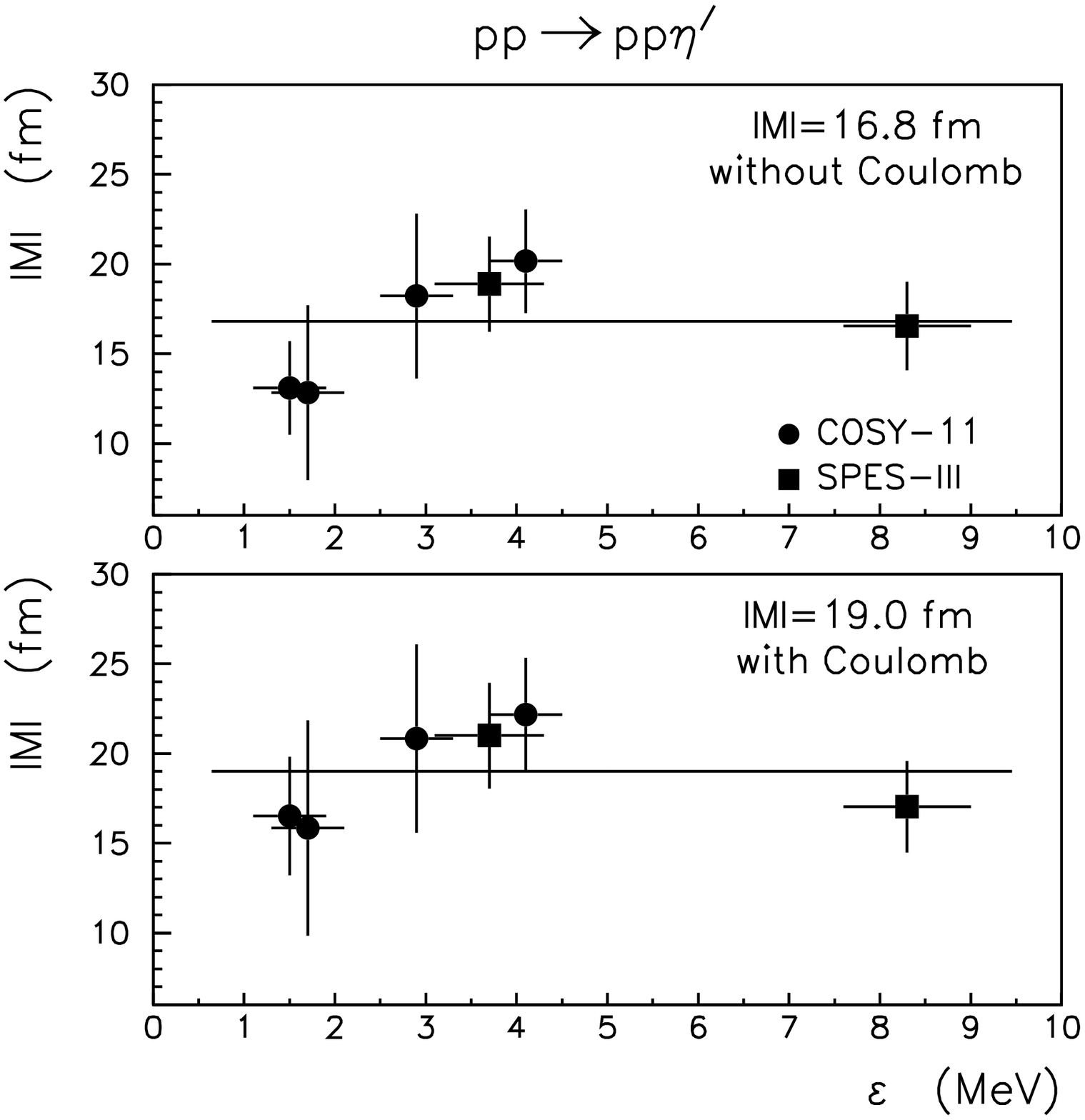,width=6.2cm,height=8cm}
\end{minipage}}
\phantom{aa}\vspace{-0.5cm}
\caption[]{Experimental data on the average 
$pp{\to}pp\eta$~\protect\cite{Bergdolt,Chiavassa,Calen1}
and  $pp{\to}pp\eta^\prime$~\protect\cite{Hibou,Moscal}
production amplitudes calculated with and without Coulomb
repulsion. The solid lines show the fit with a constant value for
the production matrix element.}
\label{memo2-1}
\phantom{aa}\vspace{-0.7cm}
\end{figure}

\begin{figure}[h]
\psfig{file=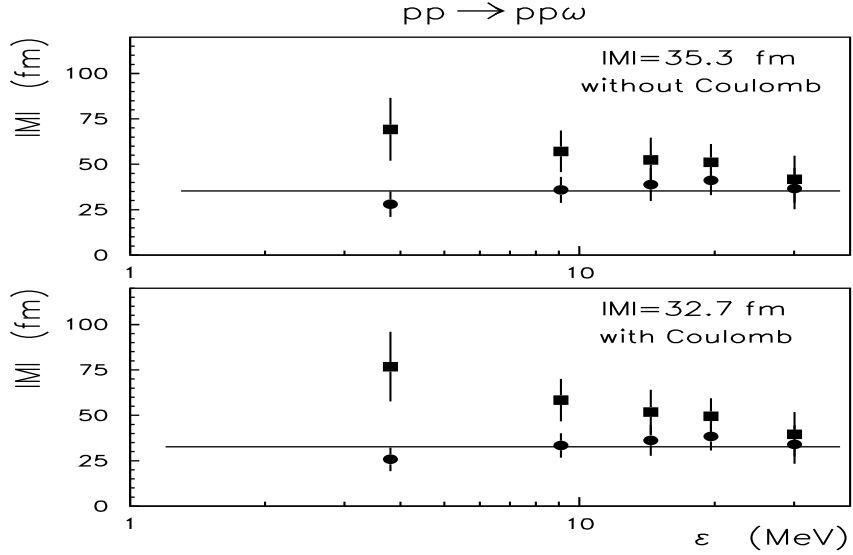,width=12.6cm,height=8cm}
\phantom{aa}\vspace{-0.8cm}
\caption[]{Experimental data on the average 
$pp{\to}pp\omega$~\protect\cite{Hibou1} production amplitudes 
calculated with and without Coulomb repulsion, for fixed 
$\omega$ mass (squares) and for the finite spectral function of the
$\omega$-meson with a width of 8.41 MeV
(circles). The solid lines show the fit with a constant value for
the production matrix element.}
\label{memo19}
\phantom{aa}\vspace{-0.5cm}
\end{figure}

In a similar way we evaluate the average production
amplitude from the total cross sections for the
$pp{\to}pp\eta$~\protect\cite{Bergdolt,Chiavassa,Calen1},
$pp{\to}pp\omega$~\cite{Hibou1}
and $pp{\to}pp\eta^\prime$~\protect\cite{Hibou,Moscal}
reactions and show the result in Figs.\ref{memo2-1},\ref{memo19}.
The results for the $pp{\to}pp\omega$ reaction are shown for 
a fixed $\omega$-meson pole mass (squares) and for the calculation 
with a Breit-Wigner $\omega$  spectral function (circles), which is
explicitely given as
\begin{eqnarray}
\label{wigner}
|M_R|=
2^{9/2} \ \pi^2 \  \lambda^{1/4}(s,m_N^2,m_N^2) \ \sqrt{\sigma s}
\left\lbrack \ \intop^{\sqrt{s}-2m_N}_{m_\pi}\hspace{-0.3cm}
\frac{\Gamma \ dx}{(x-m_\omega)^2+\Gamma^2/4} \right.
\nonumber \\
\times \left. \ \hspace{-0.2cm} \intop^{(\sqrt{s}-x)^2}_{4m_N^2}
\hspace{-0.3cm} \lambda^{1/2}(s,s_1,x^2) \ 
\lambda^{1/2}(s_1,m_N^2,m_N^2) \ \ 
\left| C_{eff}(0.5\sqrt{s_1-4m_N^2}) \right|^2 
\ \ \frac{ds_1}{s_1} \right\rbrack^{-1/2}
\end{eqnarray}
with  the vacuum $\omega$-meson width $\Gamma{=}8.41$~MeV.

Again the deviation of the matrix element $|M|$ from a constant seems
to be small for $\eta$, $\omega$ and $\eta^\prime$
production in $pp$ collisions. The data are only available 
for $\epsilon{>}1$~MeV and thus we can not observe the effect
of the Coulomb $pp$ final state repulsion. The calculations  
with and without Coulomb correction provide almost the same results
for the production amplitudes, i.e. $|M|{\approx}61$~fm
for the $\eta$, $|M|{\approx}33$~fm for the $\omega$,
and $|M|{\approx}19.0$~fm for the $\eta^\prime$-meson.
Note that in case of the $\omega$ meson it is essential to account
for the finite width of the spectral function close to threshold.

\begin{figure}[h]
\phantom{aa}\vspace{-0.5cm}
\psfig{file=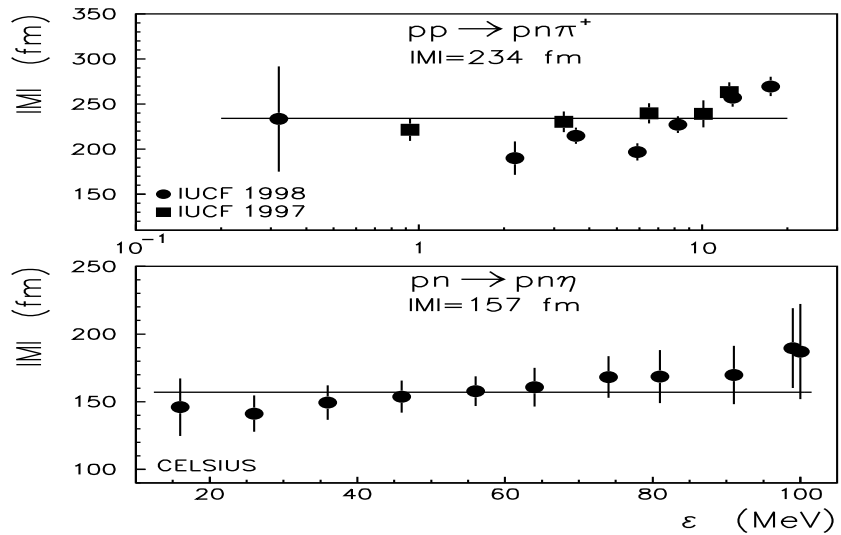,width=12.6cm,height=8cm}
\phantom{aa}\vspace{-0.72cm}
\caption[]{The data on the $pp{\to}pn\pi^+$~\protect\cite{Hardie,Flammang}
and $pn{\to}pn\eta$~\protect\cite{Calen2} production amplitudes.}
\label{memo7}
\phantom{aa}\vspace{-0.5cm}
\end{figure}

Recently IUCF published data on the $pp{\to}pn\pi^+$~ reaction 
\cite{Hardie,Flammang} 
and CELSIUS  reported $pn{\to}pn\eta$~\cite{Calen2} total cross 
sections. Both reactions are crucial for the verification of our
approach, since the final $np$ system does not suffer Coulomb
repulsion as in case of the meson production data at
$\epsilon{\le}1$~MeV. Fig.\ref{memo7} shows the $pp{\to}pn\pi^+$ 
and $pn{\to}pn\eta$ production amplitude extracted by 
Eq.\ref{average} with inclusion of the $np$ FSI. Indeed, the two
experimental points available at $\epsilon{\le}1$~MeV
as well as the data for the $pp{\to}pn\pi^+$ cross section at higher 
exess energies are reproduced by a constant value of 
$|M|\approx 234$~fm. Fig.\ref{memo7} illustrates that the data 
for the $pn{\to}pn\eta$ reaction can be reasonably described by  
$|M|\approx 157$~fm.

Finally, the simple approach outlined above allows to evaluate 
the average production amplitudes from the total cross sections
for $NN{\to}NNM$ reactions and enables one to substract
the FSI due to $NN$ rescattering. The systematical analysis
of the available experimental data on $\pi^0$, $\pi^+$,
$\eta$ and $\eta^\prime$-meson production confirms the 
validity of the method proposed. Furthermore, the results
illustrate a sensitivity to the difference between the $pp$ and
$pn$ interactions in the final state and can be  
tested by data at excess energies below 1~MeV.
Since the effective range parameters are the essential
ingredients for our calculations, the method should be
limited to $\epsilon{\leq}40$~MeV (see Fig.\ref{memo6-8}).
However, at $\epsilon{\ge}40$~MeV the $NN$ scattering amplitude 
is almost energy independent and approaches a constant value, which 
might provide an explanation for the observation that the method 
seems to work even at higher energies.

Indeed, the results for the $pp{\to}pp\eta$ and $pn{\to}pn\eta$ reactions
indicate an almost constant value of $|M|$
up to $\epsilon{\simeq}100$~MeV. This finding is in line with the
meson exchange model for $\eta$-meson production due to the
$S_{11}(1535)$ intermediate baryonic resonance excitation which
provides the dominant $S$-wave production amplitude.
A different situation holds for the $NN{\to}NN\pi$ reaction
because at large $\epsilon$ the meson exchange model involves
the ${\Delta}(1232)$ resonance and a strong contribution to the
production amplitude due to the $P$-wave. Therefore, our approach 
can not be valid for $\pi$-meson production at large $\epsilon$.

\begin{figure}[h]
\phantom{aa}\vspace{-0.7cm}
\psfig{file=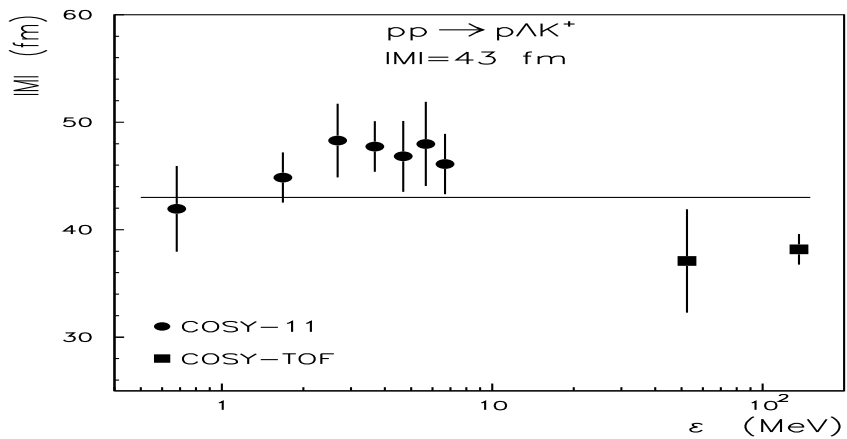,width=12.6cm,height=6.5cm}
\phantom{aa}\vspace{-0.8cm}
\caption[]{Experimental data on the average 
$pp{\to}p{\Lambda}K^+$\protect\cite{Balewski,Bilger}
production amplitudes. The solid lines show a fit 
with $|M|{=}43$~fm.}
\label{memo4}
\end{figure}

Recently the $pp{\to}p{\Lambda}K^+$ reaction was measured at 
COSY~\cite{Balewski,Bilger}. The data indicate a strong 
deviation  from the calculations with the one boson exchange 
model~\cite{Sibirtsev1,Sibirtsev4} at low $\epsilon$ due to the
FSI between the proton and $\Lambda$-hyperon~\cite{Sibirtsev2}.
We have evaluated the $pp{\to}p{\Lambda}K^+$ production amplitude 
with the singlet $^1S_0$ and triplet $^3S_1$ effective range 
parameters for ${\Lambda}p$ scattering from Ref.\cite{Stoks} 
(model a) and show the result in Fig.\ref{memo4}. Again the data can be
reasonably reproduced with $|M| \approx 43$~fm over the available range 
of the excess energy. 

We mention that the parameters for the $YN$ interactions cannot 
be fitted uniquely to the available $YN$ scattering data
since experimental results are very scarce and have large
statistical and systematical uncertainties. In turn the
$pp{\to}NYK$ reaction might serve  as an additional 
source for the examination of the hyperon-nucleon interaction at low
relative momenta. 

Furthermore, to analyze the $pp{\to}p\Sigma^0K^+$ data~\cite{Sewerin}
one needs accurate coupled channel calculations
that include the $\Sigma^0p{\leftrightarrow}{\Lambda}p$
transition as well as $\Sigma^0p{\to}\Sigma^0p$ effective 
range parameters, which are not available by now~\cite{Stoks1}.

\begin{table}[h]
\begin{center}
\phantom{aa}\vspace{-0.9cm}
\caption{\label{table1} The $pp{\to}pp\pi^0$, $pp{\to}pp\eta$,
$pp{\to}pp\omega$
and $pp{\to}pp\eta^\prime$ production amplitudes $|M|$ 
evaluated from the data with and without the Coulomb corrections.} 
\vspace{0.25cm}
\begin{tabular}{|l||c|c||c|c|}
\hline 
Reference & \multicolumn{2}{c||} {without Coulomb}   &
\multicolumn{2}{|c|} {with Coulomb}  \\
\cline{2-5} & $|M|$ (fm) & $\chi^2$ & $|M|$  (fm) & $\chi^2$ \\
\hline
\multicolumn{5}{|c|} {$pp{\to}pp\pi^0$ } \\
\hline
IUCF~~\cite{Meyer1} & 84.2 & 1.8 & 88 & 6.5 \\
IUCF~~\cite{Meyer2} & 79.0 & 0.8 & 81.7 & 4.8 \\
CELSIUS~~\cite{Bondar} & 79.9 & 7.8 & 83.1 & 13.4 \\
\hline
\hline
\multicolumn{5}{|c|} {$pp{\to}pp\eta$ } \\
\hline
SATURNE-II~~\cite{Bergdolt} & 55 & 5.1 & 62 & 3.9 \\
PINOT~~\cite{Chiavassa} & 63 & 1.9 & 60 & 3.3 \\
CELSIUS~~\cite{Calen1} & 61 & 1.1 & 61 & 2.5 \\
\hline
\hline
\multicolumn{5}{|c|} {$pp{\to}pp\omega$ } \\
\hline
SPES-III~~\cite{Hibou1} & 35.3 & 0.5 & 32.7 & 0.5 \\
\hline
\hline
\multicolumn{5}{|c|} {$pp{\to}pp\eta^\prime$ } \\
\hline
SPES-III~~\cite{Hibou} & 17.6 & 0.4 & 18.7 & 1.0 \\
COSY-11~~\cite{Moscal} & 16.1 & 1.3 & 19.3 & 0.6 \\
\hline
\end{tabular}
\end{center}
\phantom{aa}\vspace{-0.65cm}
\end{table}

The Tables \ref{table1} and \ref{table2} show the averaged 
production amplitudes evaluated from the data  
for the reactions discussed above. We separately show the 
results from different experiments, which are in reasonable 
agreement with each other. For the $pp{\to}pp\pi^0$, 
$pp{\to}pp\eta$ and $pp{\to}pp\eta^\prime$ reactions the 
results are shown with and without Coulomb correction to the $pp$ FSI.

\begin{table}[h]
\begin{center}
\phantom{aa}\vspace{-0.9cm}
\caption{\label{table2} The $pp{\to}pn\pi^+$,
$pn{\to}pn\eta$ and $pp{\to}p{\Lambda}K^+$
production amplitudes $|M|$.} 
\vspace{0.2cm}
\begin{tabular}{|l||c|c|}
\hline 
Reference &  $|M|$ (fm) & $\chi^2$  \\
\hline
\multicolumn{3}{|c|} {$pp{\to}pn\pi^+$ } \\
\hline
IUCF~~\cite{Hardie} & 240 & 1.8 \\
IUCF~~\cite{Flammang} & 228 & 6.8 \\
\hline
\hline
\multicolumn{3}{|c|} {$pn{\to}pn\eta$ } \\
\hline
CELSIUS~~\cite{Calen2}\ &  157 & 0.5 \\
\hline
\hline
\multicolumn{3}{|c|} {$pp{\to}p{\Lambda}K^+$ } \\
\hline
TOF~~\cite{Bilger} & 38 & 0.05 \\
COSY-11~~~\cite{Balewski} & 46.3 & 0.41 \\
\hline
\end{tabular}
\end{center}
\phantom{aa}\vspace{-0.8cm}
\end{table}

\begin{figure}[b]
\phantom{aa}\vspace{-0.7cm}
\centerline{
\psfig{file=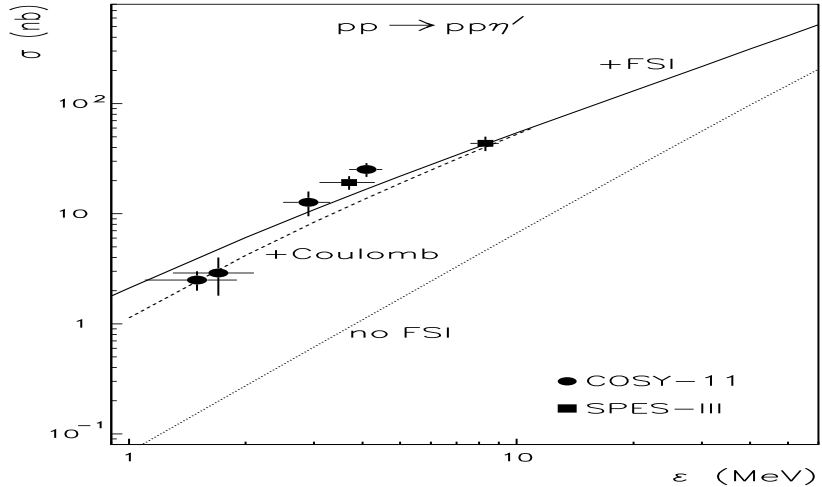,width=12.5cm,height=7cm}}
\phantom{aa}\vspace{-0.6cm}
\caption[]{The total cross section for the $pp{\to}pp\eta^\prime$
reaction. The experimental data are from 
Ref.\protect\cite{Hibou,Moscal} while the lines show the calculations
with the pion exchange model~\protect\cite{Sibirtsev5} without 
FSI between the protons (dotted), with FSI (solid) and 
with Coulomb corrections to FSI (dashed).}
\label{memo18}
\end{figure}

Finally, due to the  FSI the total production cross section is 
strongly enhanced at low excess energies as  
illustrated by Fig.\ref{memo18} which
shows the $pp{\to}pp\eta^\prime$ cross section as a function 
of $\epsilon$ calculated in the pion exchange model~\cite{Sibirtsev5}.
The dotted line indicates the calculations without the FSI
and substantially underestimates the experimental 
results~\cite{Hibou,Moscal}. Now taking into account the $s$-wave
interaction between the final protons we reasonably reproduce
the available data. Note that  the Coulomb corrections 
influence the results for $\epsilon{\le}10$~MeV. 

Our calculations illustrate  that FSI  change the energy 
dependence of the $pp{\to}pp\eta^\prime$ cross section as compared 
to the pure phase space $\epsilon^2$. Note that the results 
without FSI (dotted line in Fig.\ref{memo18}) might, in
principle,  be renormalized in order to fit the data~\cite{Hibou,Moscal}
for $\epsilon \leq$ 10 MeV, however, the increase with $\epsilon$ would
be much faster. 
This indicates that in order to determine the FSI experimentally 
one needs data on the total production cross section from threshold 
up to about 100 MeV in excess energy.

\section{FSI and  differential observables}
Obviously the FSI effect differential observables in a more
pronounced way than the total production cross section. 
Fig.\ref{memo6-8} shows that the $s$-wave dominantes the low 
energy proton-proton scattering and accordingly enhances 
the low energy part of the $pp$ invariant mass distribution.
Thus, due to energy conservation, the high energy part
of the final meson-baryon invariant mass distribution is also enhanced.
Let us illustrate this for the $pp{\to}pp\eta^\prime$  reaction.

Since there are no data on baryonic resonances that couple to the 
$\eta^\prime$-meson, our calculations~\cite{Sibirtsev5} for the
$pp \rightarrow pp \eta^{\prime}$ reaction have been carried out
within the pion exchange model without explicitly introducing 
intermediate baryonic resonances.
Thus any deviation of the calculated differential 
observables for the $pp{\to}pp\eta^\prime$ reaction at low 
$\epsilon$ from phase space only stems from 
the $s$-wave FSI between the protons.

\begin{figure}[b]
\phantom{aa}\vspace{-0.6cm}
\centerline{
\begin{minipage}[l]{6cm}
\psfig{file=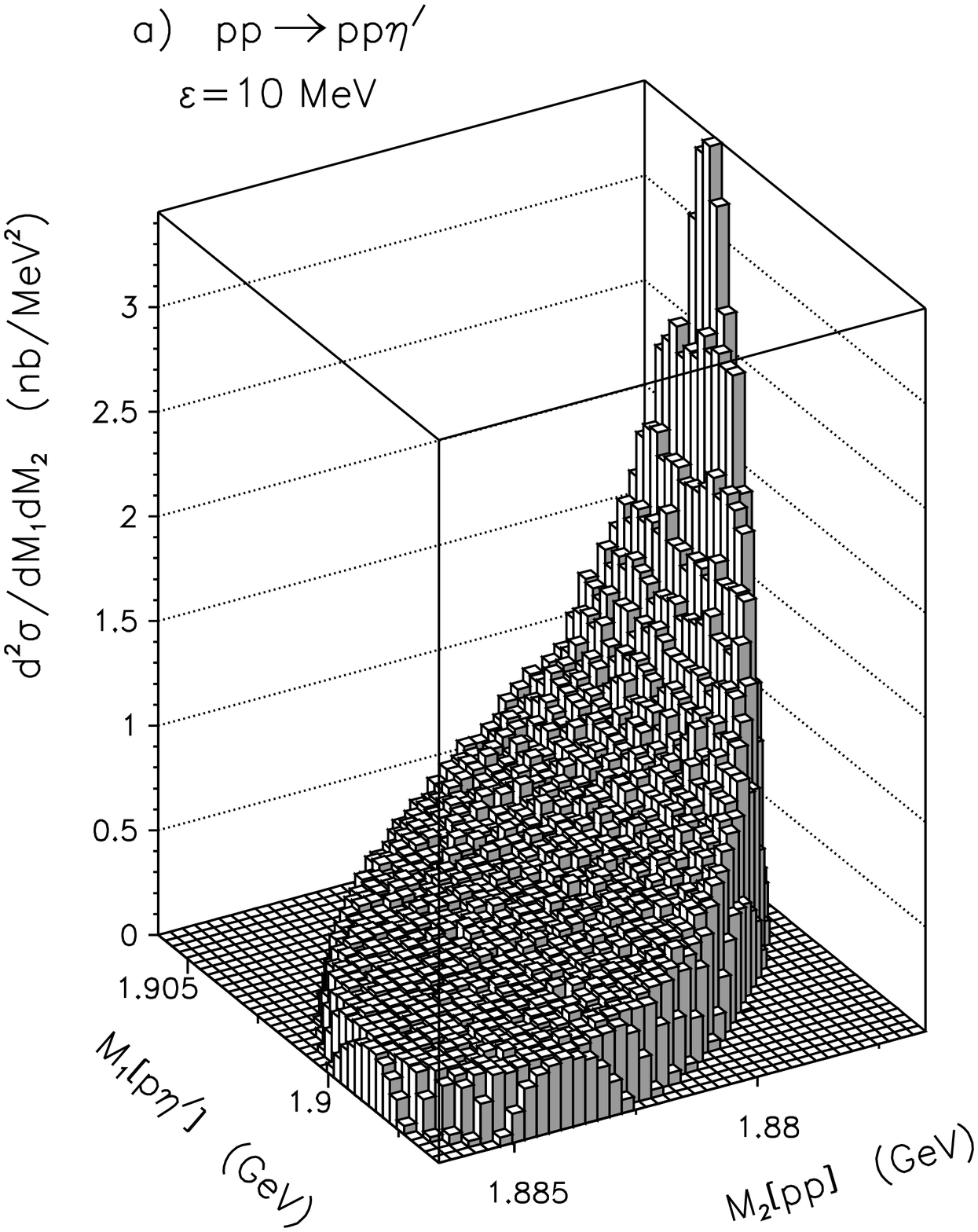,width=6.1cm,height=7.6cm}
\end{minipage}\begin{minipage}[l]{6cm}
\psfig{file=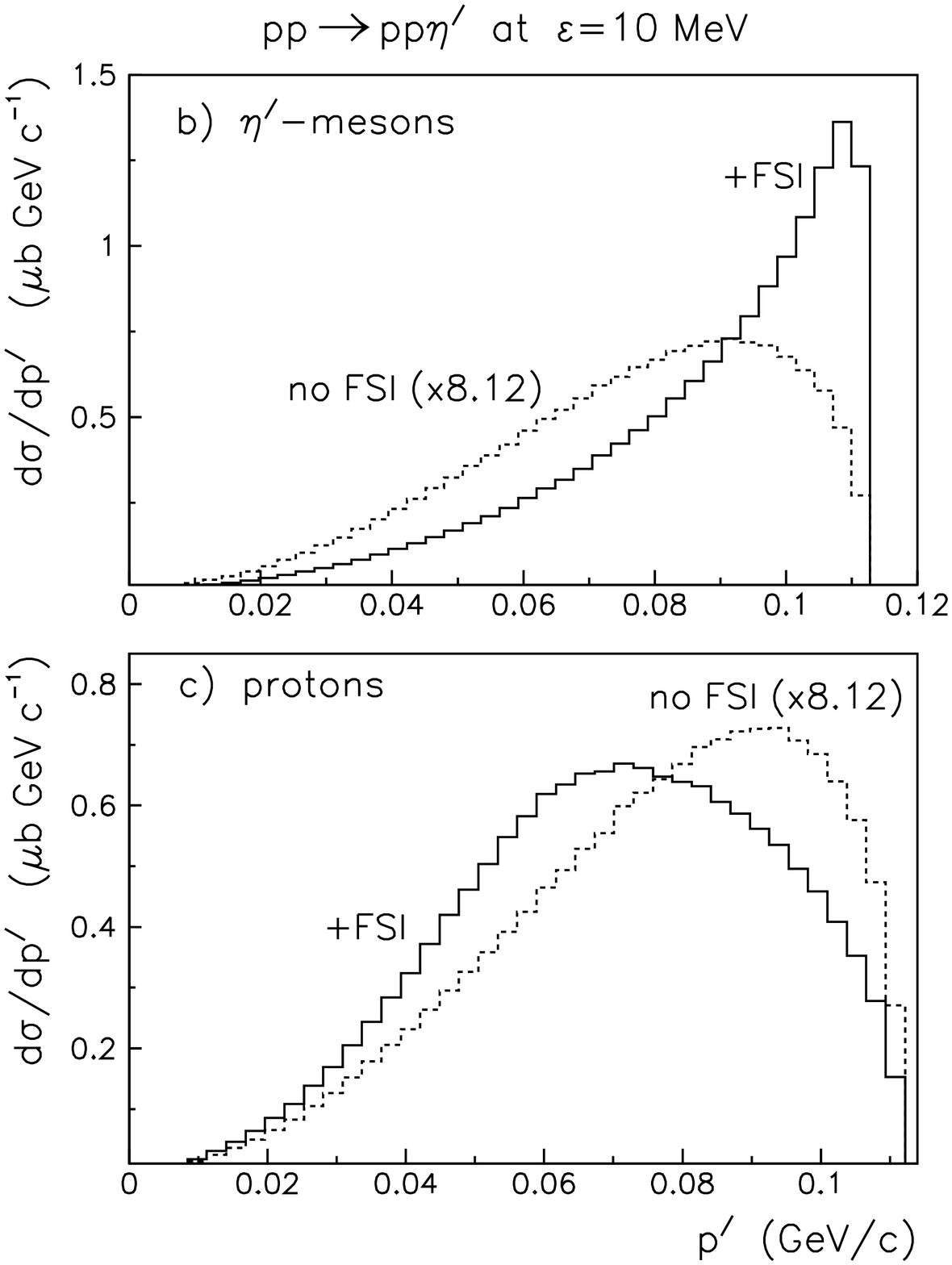,width=6.1cm,height=7.6cm}
\end{minipage}}
\phantom{aa}\vspace{-0.6cm}
\caption[]{The differential observables for the
$pp{\to}pp\eta^\prime$ reaction at $\epsilon{=}10$~MeV. 
a). The Dalitz plot calculated within the pion exchange model and
with FSI. b-c). The momentum spectra of the $\eta^\prime$-mesons
and protons in the center of mass system calculated with
(solid) and without FSI (dashed histograms). The calculations 
without FSI have been renormalised to the same total cross section.}
\label{memo12-13}
\end{figure}

Fig.\ref{memo12-13}a) shows the Dalitz plot for the
$pp{\to}pp\eta^\prime$ reaction at $\epsilon{=}10$~MeV. 
Indeed the distribution is enhanced at low $pp$ and large
$p\eta^\prime$ masses.  Figs. \ref{memo12-13}b,c), furthermore, show
the c.m.s. momentum spectra of the $\eta^\prime$-mesons
and protons produced in the $pp{\to}pp\eta^\prime$ reaction 
at $\epsilon{=}10$~MeV. The solid histograms display our calculations
within the pion exchange model~\cite{Sibirtsev5} including the FSI. The 
dashed histograms are the results without FSI but corrected by a 
factor $8.12$ due to the difference in the total cross section 
calculated with and without FSI (see Fig.\ref{memo18}).
The impact of the FSI is obvious and can be easily detected
in the $\eta^\prime$-spectra. It is important to note that
the distortion of the phase space distribution due to the FSI
should be properly taken into account when extrapolating experimental 
data in a limited acceptance to $4\pi$.

\begin{figure}[h]
\centerline{
\psfig{file=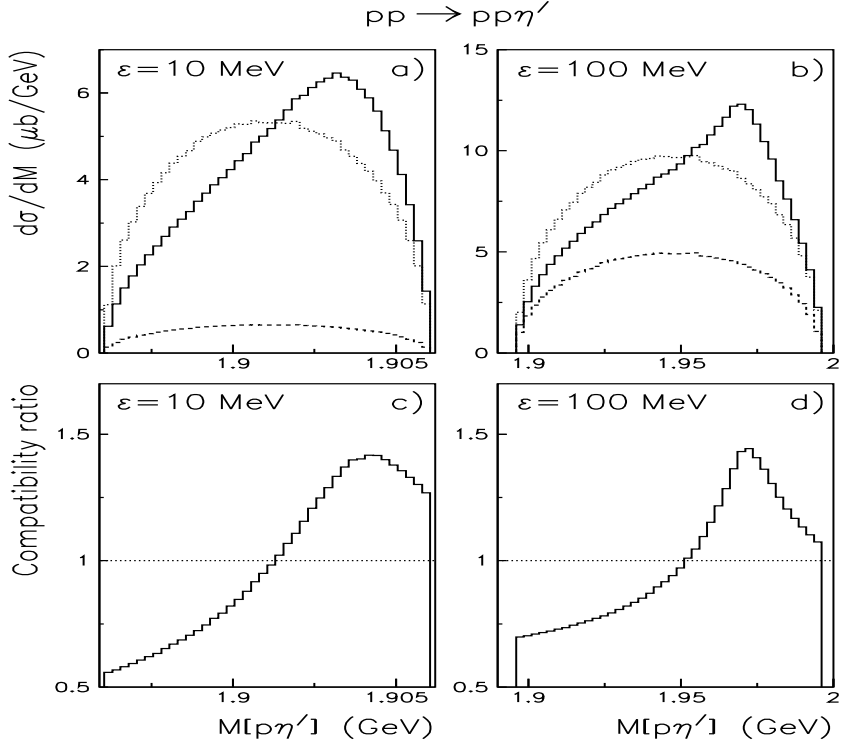,width=12.5cm,height=10cm}}
\phantom{aa}\vspace{-0.6cm}
\caption[]{The $p\eta^\prime$ invariant mass spectra (a,b)
and the compatibility ratio (c,d) calculated for the 
$pp{\to}pp\eta^\prime$ reaction at $\epsilon{=}10$~MeV and 100~MeV.
The histograms in a),b) show the results with FSI (solid),
without FSI (dotted) and without FSI but renormalized to the same
cross section(dashed).}
\label{memo11}
\phantom{aa}\vspace{-0.8cm}
\end{figure}

Moreover, the FSI produce some resonance structure in the
meson-baryon invariant mass distribution as shown in
Fig.\ref{memo11}a,b) for the $pp{\to}pp\eta^\prime$ reaction at 
$\epsilon{=}10$~MeV and $\epsilon{=}100$~MeV. Here the solid
histograms are our calculations with FSI while the dotted histograms
show the results calculated without FSI which are similar to
the pure phase-space distributions.  The dashed histograms 
in Figs. \ref{memo11}a,b) are the calculations 
without FSI but renormalized to the same total production cross
section. Recall that we do not include intermediate 
baryonic resonances in our model~\cite{Sibirtsev5} and that the
$pseudo$ resonance structure in the $p\eta^\prime$ mass
spectra stems from the FSI.

Experimentally this effect can be detected when analyzing the
compatibility ratio, i.e. the ratio
of the measured invariant mass spectra to the phase space 
distribution that is normalized to the experimental total cross section.
The calculated compatibility ratio for the $pp{\to}pp\eta^\prime$
reaction at $\epsilon{=}10$~MeV and 100~MeV is shown in 
Figs. \ref{memo11}c,d)
and visibly deviates from unity. Recall that in the absence of
FSI as well as other effects, e.g. an excitation of a 
baryonic resonance in the meson-baryon system or the appearence 
of higher partial waves in the production amplitude 
(which might happen at large $\epsilon$), the compatibility  ratio
should approach unity. On the other hand, to detect the distortion
of the compatibility ratio one needs sufficiently large statistical 
accuracy as can be seen from Figs. \ref{memo11}c,d).

\begin{figure}[h]
\centerline{
\psfig{file=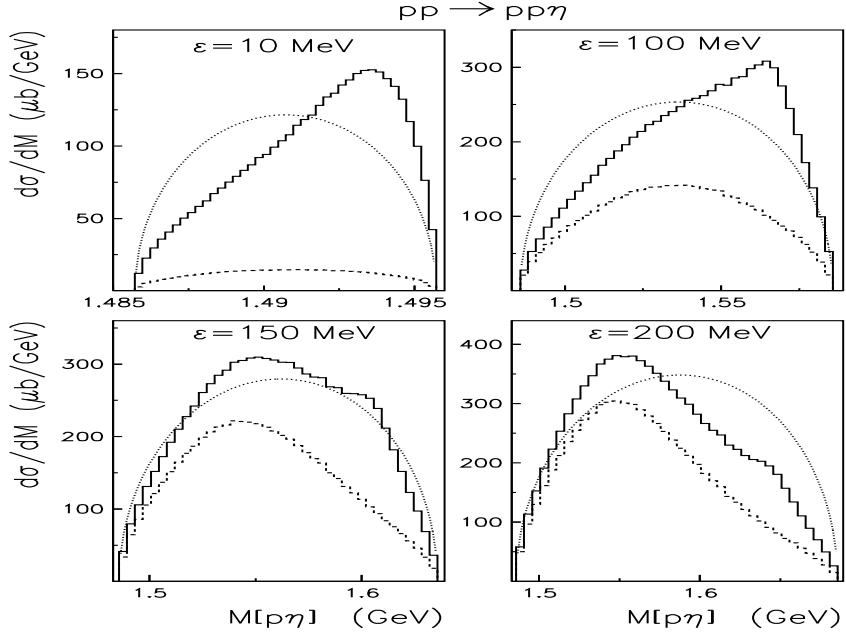,width=12.5cm,height=9cm}}
\phantom{aa}\vspace{-0.6cm}
\caption[]{The $p\eta$ invariant mass spectra
for the $pp{\to}pp\eta$ reaction at $\epsilon{=}10$~MeV,
100, 150 and  200~MeV. The solid histograms are calculations with 
an excitation of $S_{11}(1535)$ and FSI, while the dashed histograms
show the results with $S_{11}(1535)$ but without FSI. The
dotted lines indicate the normalized phase-space distributions.}
\label{memo14}
\phantom{aa}\vspace{-0.7cm}
\end{figure}

In order demonstrate how an intermediate resonance shows up in the
invariant mass spectra we
analyze the $pp{\to}pp\eta$ reaction calculating the 
production amplitude due to the excitation of the $S_{11}(1535)$
resonance. Fig.\ref{memo14} shows the resulting $p\eta$ invariant
mass spectra for $\epsilon{=}10$, 100, 150 and 200~MeV.
The solid histograms are our calculations with the $S_{11}(1535)$ and
FSI, while the dotted histograms indicate the results without
FSI between the protons. The dotted lines in Fig.\ref{memo14}
show the phase-space distribution normalized to the calculated
total cross section.

As discussed above, the $S_{11}$ structure cannot be detected at
$\epsilon{\le}100$~MeV since the width of the baryonic resonance is
larger than the range of the $p\eta$ invariant mass. Furthermore,
the shape of the spectra calculated with a $S_{11}$ intermediate resonance
and without FSI are similar to the spectra in line with phase space
at $\epsilon{\le}100$~MeV. The deviation of the $p\eta$ mass
spectra at $\epsilon{=}10$ and 100~MeV from phase space
(dotted lines) is entirely due to FSI.

The $S_{11}$ structure can be detected at $\epsilon{=}150$ and
200~MeV where the $p\eta$ mass spectra calculated even without FSI 
(dashed histograms) differ already from pure phase space. 
Note, however, that FSI substantially
distort the spectra and consequently we find two structures in the 
$p\eta$ invariant mass distributions. The enhancement around
$M_{p\eta}$ is due to the $S_{11}$ resonance while the
structure close to the kinematical limit of the $p\eta$ 
mass spectra stems from the FSI. Again the compatibility ratio 
might serve as a promising tool to detect the reaction  mechanism.

\begin{figure}[h]
\centerline{
\psfig{file=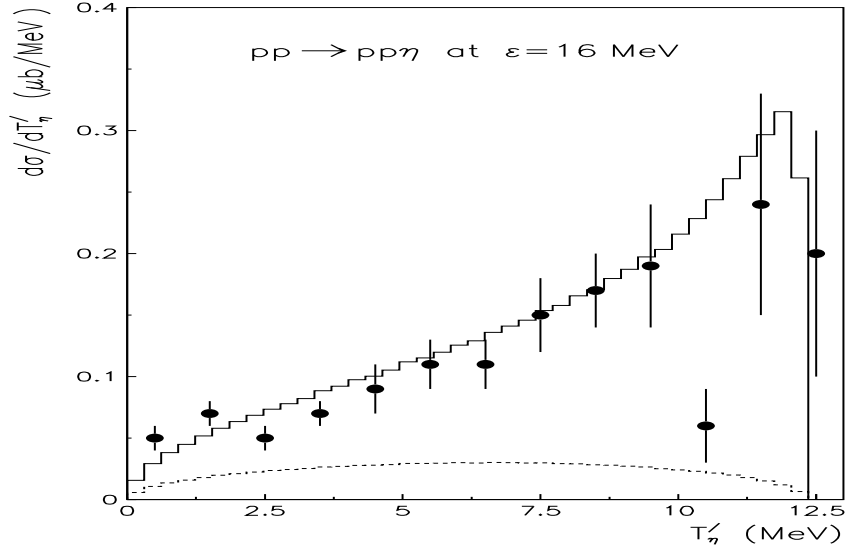,width=12.5cm,height=8cm}}
\phantom{aa}\vspace{-0.6cm}
\caption[]{The $\eta$-meson energy spectrum in the
center-of-mass system measured for the $pp{\to}pp\eta$ reaction 
at $\epsilon{=}16$~MeV. The full dots show the experimental results from 
Ref.\protect\cite{Calen4} while the solid histogram is our calculation 
with FSI, the dashed histogram without FSI.}
\label{memo15}
\phantom{aa}\vspace{-0.7cm}
\end{figure}

Recently CELSIUS reported~\cite{Calen4} the $\eta$-meson 
c.m.s. energy spectrum measured in the $pp{\to}pp\eta$ reaction 
at $\epsilon{=}16$~MeV which is shown in Fig.\ref{memo15}
together with our calculations. The solid histogram in
Fig.\ref{memo15} shows the result with FSI that
reasonably reproduces the data; the dashed histogram indicates 
the result without FSI and substantially differs from 
the experimental spectrum both in the absolute height
and in shape. This comparison, furthermore, demonstrates
the validity of our approach which is of sufficient simple form to be
used in all data analysis for near threshold reactions.

\section{Summary}
In this work we have proposed a simple method to analyze or
calculate cross sections on near threshold meson production in 
$pp$ collisions by dividing out kinematical factors and accounting for 
final-state-interactions (FSI) between the nucleons including 
approximately also Coulomb corrections. Our analysis of the various 
models for FSI has shown that the inverse Jost-function method has 
the largest range of applicability, posesses the correct boundary 
condition for large excess energies and, furthermore,
only involves the effective range parameters $a_s$ and $r_s$ that can be
taken from a fit to the respective $s$-wave scattering amplitude. 
 
Within this model we have analyzed the available data on $\pi$, 
$\eta$, $\omega$, $\eta^\prime$ and $K^+\Lambda$ production and found 
that all data are approximately compatible with constant production 
matrix elements. This information now in  turn can be used to calculate 
reaction channels with different final states of the baryons if their 
FSI is known. On the other hand, the constant matrix
element hypothesis allows to $measure$ the FSI of baryons that are not 
available for scattering experiments. Note, however, that precise data up to
excess energies of $\approx$ 100 MeV will be necessary.

Furthermore, we have shown that a differential data analysis in terms of 
Dalitz-plots allows to distinguish effects from final state interactions
and resonance amplitudes if data are available in a sufficiently wide
energy range comparable at least to the width of the resonance amplitude.

\end{document}